\documentclass[prb,twocolumn,aps,showpacs,superscriptaddress]{revtex4-1}
\usepackage{color}
\usepackage{amsmath}
\usepackage{graphicx}

\begin{document}

\title{The effect of spin-orbit interaction and attractive Coulomb potential on the magnetic properties 
of Ga$_{1-x}$Mn$_{x}$As}

\author{A.-M. Nili}
\address{Department of Physics and Astronomy \& Center for Computation and
Technology, Louisiana State University, Baton Rouge, Louisiana 70803}
\author{M. A. Majidi}
\address{Departemen Fisika, FMIPA, Universitas Indonesia, Depok 16424, 
Indonesia}
\author{P. Reis} 
\author{J. Moreno}
\author{M. Jarrell}
\address{Department of Physics and Astronomy \& Center for Computation and
Technology, Louisiana State University, Baton Rouge, Louisiana 70803}

\date{\today}

\begin{abstract}  
We employ the 
dynamical mean-field approximation to study the magnetic properties of a model relevant for the 
dilute magnetic semiconductors. 
Our model includes the spin-orbit coupling on the hole bands, the exchange interaction, 
and the attractive Coulomb potential between the negatively charged magnetic ions and the itinerant holes.
The inclusion of the Coulomb potential 
significantly renormalizes the exchange coupling and
enhances the ferromagnetic transition temperature for a wide range 
of couplings.
We also explore the effect of the spin-orbit interaction by using two different 
values of the ratio of the effective masses of the heavy and light holes.
We show that in the regime of small 
$J_{c}$-$V$ the spin-orbit interaction enhances $T_{c}$, while for large enough values of $J_{c}$-$V$
magnetic frustration reduces $T_c$ to values comparable to the previously calculated strong coupling limit.
\end{abstract}

\maketitle
\section{Introduction}

Although the notion of using magnetic semiconductors in spintronic devices dates back to 
the 1960's \cite{p_baltzer_66a}, the discovery 
of high temperature 
ferromagnetism in dilute magnetic semiconductors (DMS)\cite{Ohno96,mun:89} 
initiated an active search for the optimal compound with a magnetic transition above room 
temperature.
Since these materials are good sources of polarized charge carriers, they may form the basis of future 
spintronic devices,\cite{i_zutic_04,s_wolf_01} which utilize the spin of the carriers as well as their charge
to simultaneously store and process data. 
Perhaps one of the most promising DMS is GaAs doped with Manganese due to 
its rather high ferromagnetic transition temperature ($T_{c}>150$ K for bulk samples and $\sim$ 250 K 
for $\delta$-doped heterostructures\cite{a_macdonald_05a,a_nazmul_05a}) 
and its wide use in today's electronic devices.

In Ga$_{1-x}$Mn$_{x}$As, the Mn$^{+2}$ ion primarily replace Ga$^{+3}$ playing the role of acceptor by 
introducing an itinerant hole to the p-like  valence band. The strong spin-orbit interaction 
in the valence band couples 
the angular momentum to the spin of the itinerant hole resulting in total spin 
$J$=$l+s$=3/2 for the two upper valence bands and $J$=$l-s$=1/2 for the split-off band. 
Each manganese also introduces a localized spin ($S$=5/2) due to its half-filled 
$d$ orbital. In addition, since the Mn$^{+2}$ ion is negatively charge with respect to the Ga$^{+3}$ ionic 
background there is an effective attractive interaction
between the Mn ion and the charge carriers. 

In previous studies\cite{k_aryanpour_05a,m_majidi_06a} some of us  
have explored the effect of the strong spin-orbit coupling on the 
ferromagnetic transition temperature $T_{c}$, the carrier polarization as well as the density of states 
and spectral functions using the Dynamical Mean-Field Approximation (DMFA). In these studies 
we used the  $k\cdot p$ Hamiltonian to model the dispersion of the parent material (GaAs). 
While $k\cdot p$ is a good approximation around the center of the Brillouin zone ($\Gamma$ point), 
it is a poor one away from it. In this work we improve our model by incorporating a more realistic 
tight binding dispersion for the valence bands as well as an attractive on-site potential between the 
Mn ions and the itinerant holes. Moreover, we study the effect of the spin-orbit interaction of the holes 
on the magnetic behavior of the DMS. We find that  
for intermediate values of the exchange coupling both the on-site potential and the spin-orbit enhances the 
critical temperature, while in the strong coupling regime 
the spin-orbit interaction significantly suppresses $T_{c}$\cite{j_moreno_06a}. 

The effect of the attractive Coulomb potential has been discussed for models with only one valence band,
which ignore the spin-orbit interaction,\cite{m_takahashi_03a,f_popescu_07a,m_calderon_02,e_hwang_05} 
and multi-band tight-binding models, which include spin-orbit coupling, but with a  limited sampling 
of disorder configurations.\cite{a_moreo_07}
Here we include on an equal footing the effect of the attractive 
Coulomb potential using a simple Hartree term, the exchange between magnetic ions and itinerant holes, 
the spin-orbit coupling, and the disorder within the coherent potential approximation 
(CPA).\cite{d_taylor_67,p_soven_67,leath66}
We investigate the 
ferromagnetic transition temperature, the average magnetization of the Mn ions, the 
polarization of the holes,  
and the quasiparticle density of states as function of the Coulomb  
and exchange couplings. 
First, we use a single band model where spin-orbit interaction 
is ignored and carriers have angular momenta $J= 1/2$.
Next, we introduce the spin-orbit coupling in a two-band model with  $J= 3/2$. 
By changing the ratio of the masses of 
the light and heavy bands ($m_{l}$/$m_{h}$) we explore the effect of spin-orbit coupling.
This is the minimal model that qualitatively captures the physics of DMS, 
however, a more realistic approach should incorporate the conduction and split-off bands 
and this will be discuss in future studies.

\section{Model}

\begin{figure}[t]
\begin{centering}
\includegraphics[width=3.2in]{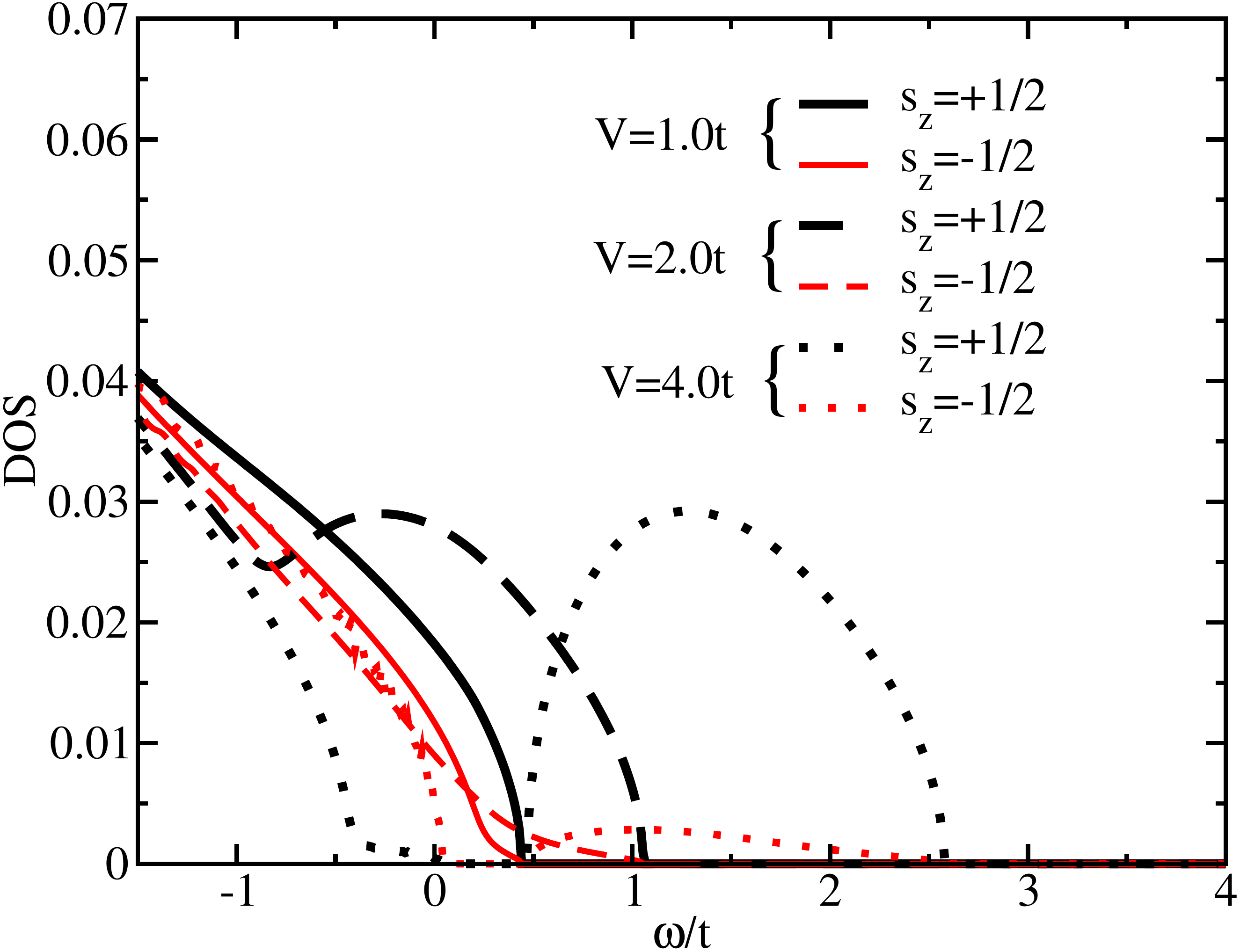}
\caption{(color online). Spin-dependent density of states for temperature T=0.01t, exchange coupling $J_{c}$=$2\,t$ 
and Coulomb potential $V=1\,t, 2\,t$ and $4\,t$. T=0.01t is below the ferromagnetic transition for all
values of $V$.} 
\label{DOS_J2}
\end{centering}
\end{figure}

We employ the simplified Hamiltonian proposed by $\rm{Zar\acute{a}nd}$ and $\rm{Jank\acute{o}}$\cite{g_zarand_02a}
with an additional Coulomb potential term:
\begin{equation}
H= H_{0} + J_{c} \sum_{i} \mathbf{S}(R_i) \cdot \mathbf{J}(R_i) +V\sum_{i} n(R_i) ,
\label{hamiltonian}
\end{equation}
where $H_{0}$ includes both electronic dispersion and spin-orbit coupling of the holes in the parent compound,
$J_{c}$ is the exchange coupling, $V$ the Coulomb strength,  $\mathbf{S}(R_i)$, $\mathbf{J}$($R_{i}$) 
and $n(R_i)$ are, respectively, the spin of the localized moment, the total angular momentum density 
and the density of the carriers at random site $i$. Short range direct or superexchange between Mn ions 
is ignored since we are in the dilute limit and we are not including clustering effects.

As discussed previously,\cite{k_aryanpour_05a,m_majidi_06a} within the DMFA 
the coarse-grained Green function matrix is: 
\begin{equation}
\hat{G}(i\omega_{n}) = \frac{1}{N} \sum_{k}[i\omega_{n}\hat{I}-\hat{H}_{0}(k)+
\mu\hat{I}-\hat{\Sigma}(i\omega_{n})]^{-1},
\end{equation}
where $N$ is the number of $k$ points in the first Brillouin zone, $\mu$ the chemical potential, and
$\hat{H}_{0}(k)$ and $\hat{\Sigma}(i\omega_{n})$, are matrices representing the band structure of the parent
material and the selfenergy, respectively.
The mean field function  $\hat{{\cal G}_{0}}(i\omega_{n}) = [\hat{G}^{-1}(i\omega_{n}) + 
\hat{\Sigma}(i\omega_{n})]^{-1}$ 
is required to solve the DMFA impurity problem. At a non-magnetic site, the Green function is simply the mean field 
function $\hat{G}_{non}(i\omega_{n})=\hat{{\cal G}_{0}} (i\omega_{n})$. The Green function at a magnetic site is
$\displaystyle{ \hat{G}_{\mathbf S}(i\omega_{n})= [{\cal \hat{G}}_{0}^{-1}(i\omega_{n})+J_{c} {\mathbf S} \cdot \hat{\mathbf J}+V]^{-1}}$
for a given local spin configuration. 

Next we average $\hat{G}_{mag}$ over different spin orientation of 
the local moment. The relatively large magnitude of the Mn moment justifies a classical treatment 
of its spin. To get the average over the angular distribution we use the effective action 
\cite{n_furukawa_98a,n_furukawa_94a}

\begin{equation}
 S_{eff}(\mathbf S)=-\sum_{n} \log\det[\hat{\cal G}_{0}^{-1}(i\omega_{n}) +
J_{c} {\mathbf S} \cdot \hat{\mathbf J} +V]e^{i\omega_{n}0^{+}}.
\end{equation}
The average over spin configuration is 
\begin{equation}
\langle\hat{G}_{mg}(i\omega_{n})\rangle= 
\frac{1}{Z}\int d\Omega_{\rm{S}}\hat{G}_{\mathbf S}(i\omega_{n})\rm{exp[-S_{eff}(\mathbf S)]} ,
\end{equation}
where $Z$ is the partition function, $Z=\int d\Omega_{\rm\textbf{s}}\rm{exp}(-S_{eff}(\textbf{S}))$. Finally the 
disorder is treated in a fashion similar to the coherent phase approximation 
(CPA)\cite{d_taylor_67,p_soven_67,leath66} and the averaged Green function reads 
$\hat{G}_{avg}(i\omega_{n})=\langle\hat{G}_{mg}\rangle x+{\cal \hat{G}}_{0}(i\omega_{n})(1-x)$ where $x$ is the doping. 

\begin{figure}[t]
\begin{center}
\includegraphics[width=.45\textwidth]{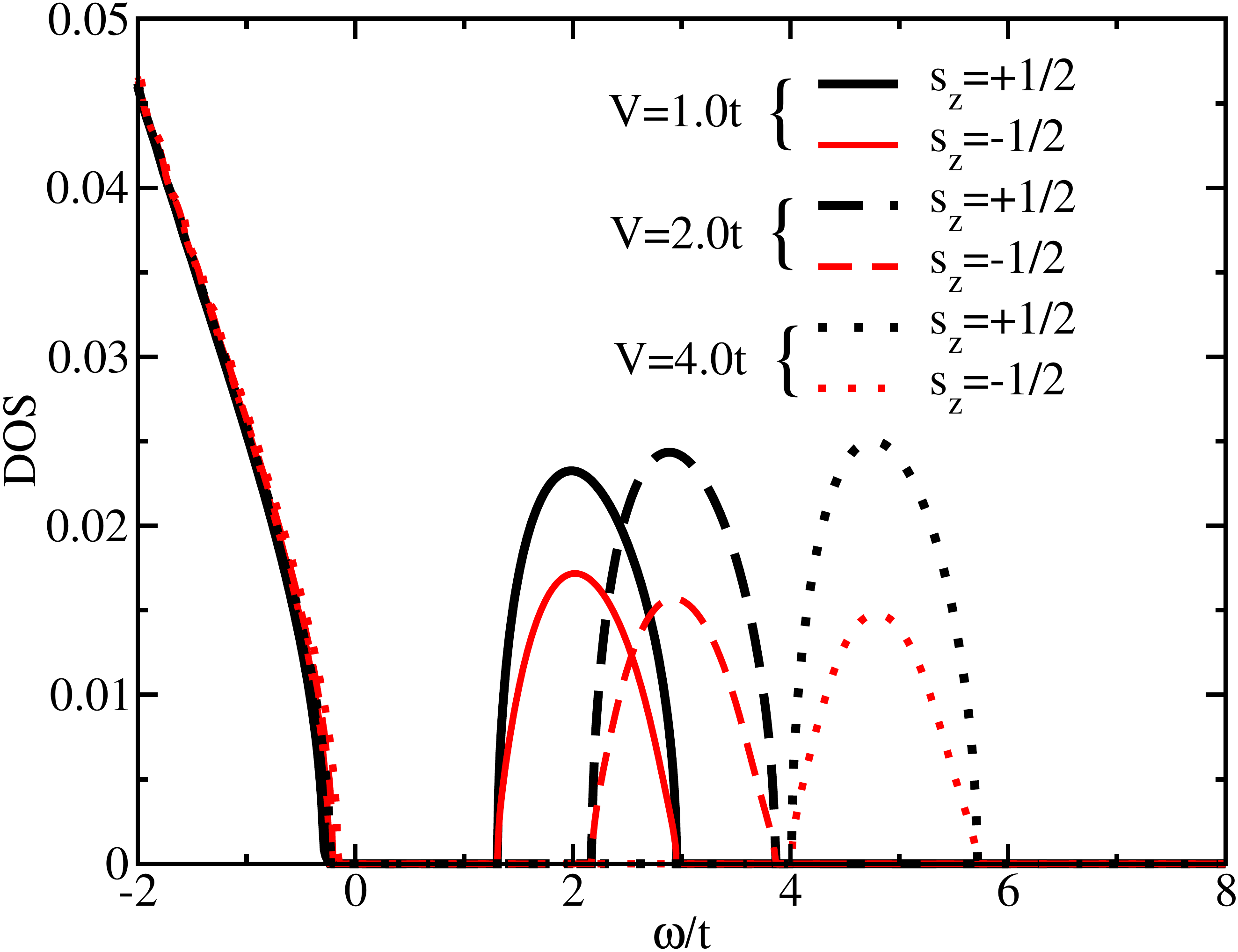}
\caption{(color online). Spin-dependent density of states for temperature T=0.04t, exchange coupling $J_{c}$=$5\,t$ 
and various values of the Coulomb potential coupling $V$. }
\label{DOS_J5}
\end{center}
\end{figure}

We obtain the hole density of states from the coarse-grained Green function in real frequency domain:

\begin{equation}
 \hat{G}(\Omega)=\frac{1}{N}\sum_{k}[\Omega\hat{I}-\hat{H}_{0}(k)-\hat{\Sigma}(\Omega)]^{-1} 
\end{equation}
where $\Omega=\omega+i0^{+}$.\
The total density of states (DOS) is
\begin{equation}
 DOS(\Omega)= -\frac{1}{\pi}\rm{Im Tr \hat{G}(\Omega)} ,
\end{equation}
where Tr is the trace. Each diagonal element of the Green function ($\displaystyle{-\frac{1}{\pi}Im \hat{G}(\Omega)}$) 
corresponds to the density of states for a specific $J_{z}$ component.\\

\section{Results}

\begin{figure}[t]
\begin{center}
\includegraphics[width=3.2in]{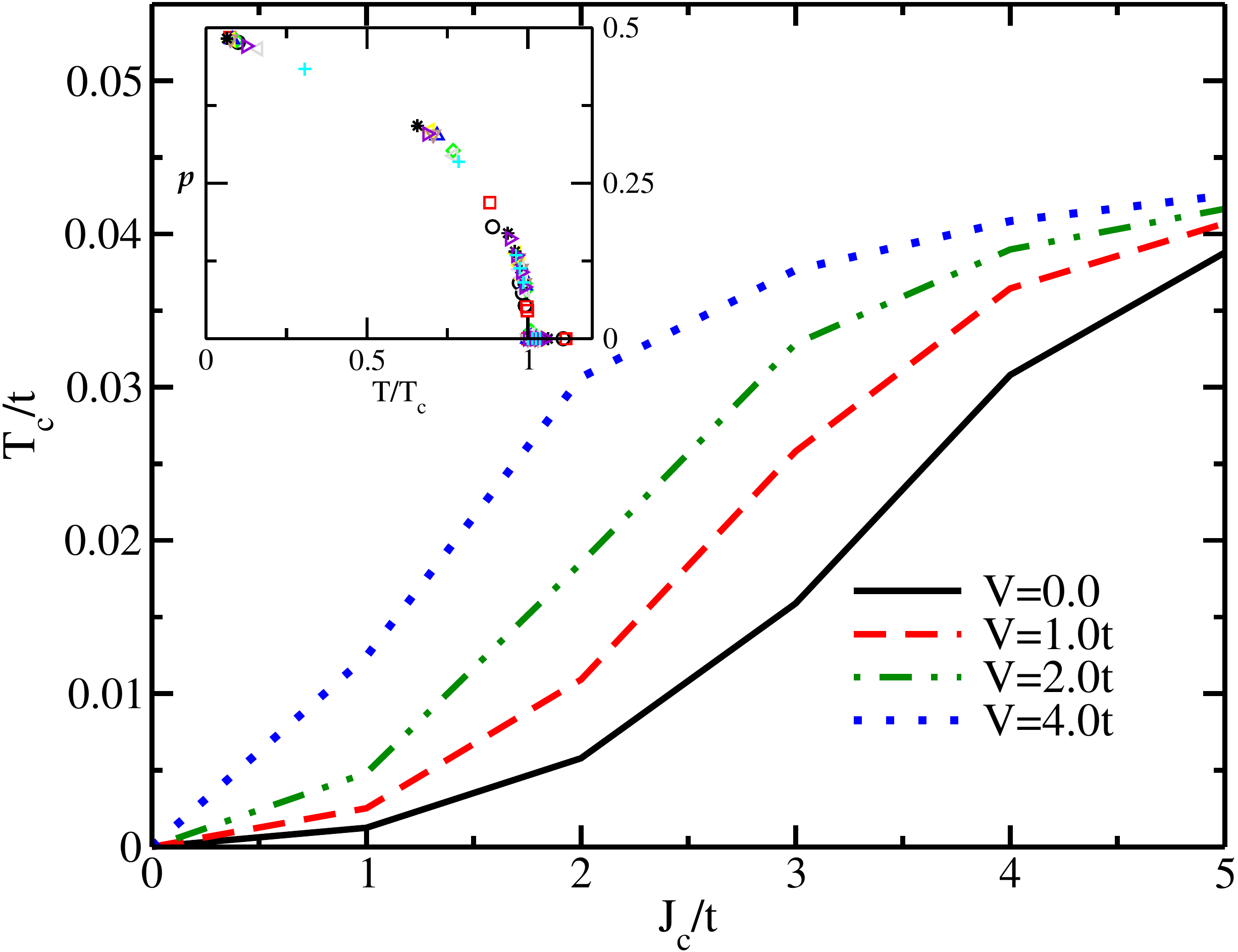}
\caption{(color online). Ferromagnetic transition temperature versus magnetic exchange coupling for
various values of the Coulomb potential. Inset: polarization of the holes as function of $T/T_c$
for a wide range of values of $J_c$ and $V$. Notice that all the polarization data collapse on 
a single curve.}
\label{Tc_vs_J}
\end{center}
\end{figure}

Since Ga$_{1-x}$Mn$_{x}$As is grown using out of equilibrium techniques a
noticeable fraction of manganese lies not on the Ga site (substitutional) but on the As site (anti-site) 
or somewhere in the middle of the crystal structure (interstitial)\cite{t_jungwirth_06}.
The real nature of interstitial defects is still controversial and yet to be resolved, \cite{j_masek_04,j_blinowski_03} 
but the one consensus is that in most samples there is strong 
compensation of the holes introduced by substitutional Mn. The density of carriers can also be controlled 
with electric fields.\cite{Ohno00} We take these considerations into account by 
simply setting the filling of the holes to half of the nominal doping\cite{j_moreno_06a}.
We focus on the doping $x$=5\% and hole filling  of $n_h=x/2$.

\begin{figure}[b]
\begin{centering}
\includegraphics[width=0.45\textwidth]{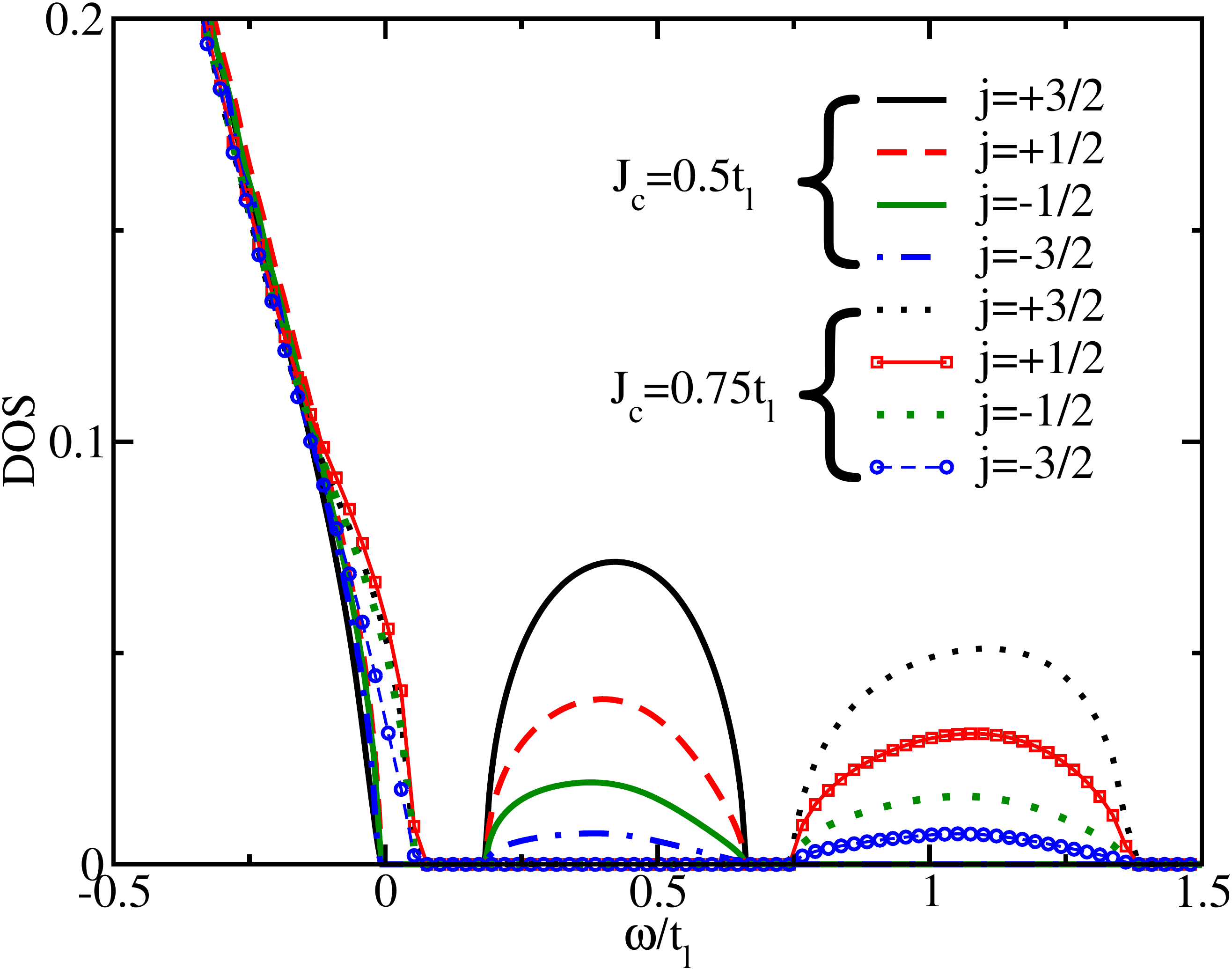} 
\caption{(color online). Density of states for $V$=0. and $J_{c}=$0.5\,t$_{l}$, 
T=0.005\,t$_{l}$; and $J_{c}=$0.75\,$t_{l}$ at a 
temperature T=1/130\,t$_{l}$. 
The impurity band is well formed with $J_{c}$=0.5$t_{l}$ and increasing of the coupling 
shifts the impurity band to higher energies.}
\label{DOS_V0}
\end{centering}
\end{figure}

\begin{figure}[b]
\begin{centering}
\includegraphics[width=0.45\textwidth]{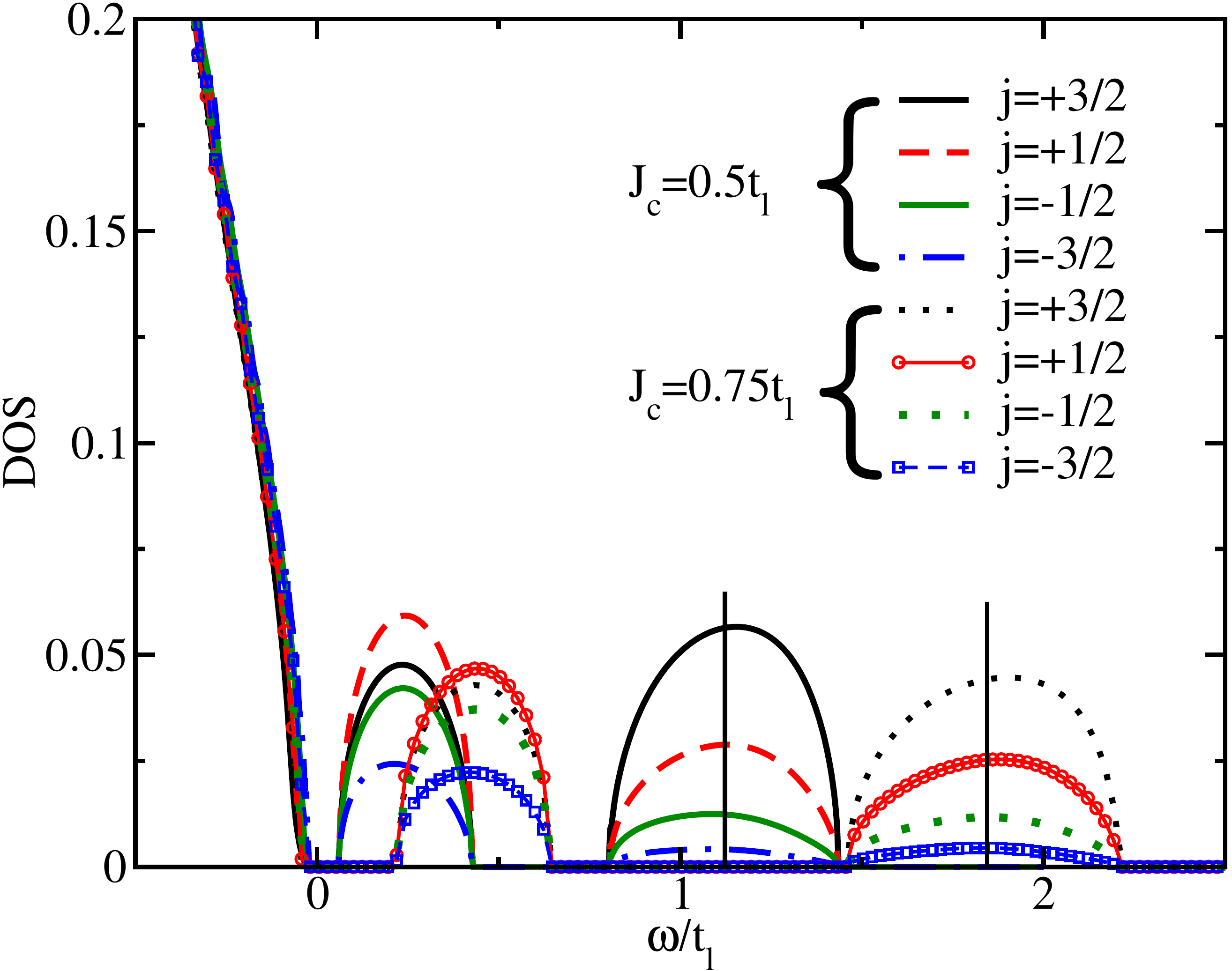} 
\caption{(color online). Density of states for $V$=\,$t_{l}$, $J_{c}$=0.5\,$t_{l}$ 
T=0.005\,t$_{l}$; and $J_{c}=$0.75\,$t_{l}$ at a 
temperature T=1/130\,t$_{l}$. 
The attractive potential enhances the formation of the impurity band as compared with Fig. \ref{DOS_V0}.
The chemical potential lies in the middle of the first impurity band, as it is displayed by the vertical
black line.}
\label{DOS_V1}
\end{centering}
\end{figure}

We start by discussing a simplified one-band model where we ignore the spin-orbit interaction. 
Our carrier dispersion is $\epsilon_{k}= -2 t (\rm{cos}(k_x) + \rm{cos}(k_y) + \rm{cos}(k_z))$, 
where $t$ is the spin independent hopping integral.
Fig. \ref{DOS_J2} and \ref{DOS_J5} display the spin-dependent density 
of states (DOS) close to the edge of the valence band for coupling constant  $J_{c}$=$2\,t$ 
and $5\,t$, respectively. Note that inclusion of the 
spin-independent attractive potential results in shifting the energy of the holes (electrons) to lower (higher) 
energies for both spin species. This is in agreement with previous 
studies \cite{m_takahashi_03a,f_popescu_07a}. 
Fig.~\ref{DOS_J2} illustrates the strong influence of the Hartee term on the states close to the valence band
edge for moderate exchange coupling. It is clear that increasing the Coulomb potential accelerates the 
formation of the impurity band and its splitting from the valence band. 
Fig.~\ref{DOS_J5}  shows that for couplings as large as $J_{c}$=5\,$t$ 
the impurity band is well formed even for relatively small Coulomb potentials ($V$=1\,$t$) and  the mere effect 
of the Coulomb term is to shift the impurity band. Notice also that the predicted shift of the 
impurity band is too large. We believe that this is a consequence of 
excluding the conduction band from our model, since band repulsion
with the conduction band pushes the impurity band to lower energies. 

The main panel in 
Fig.~\ref{Tc_vs_J} shows the dependency of $T_{c}$  on the exchange coupling  for different Coulomb potentials
within this simplified one-band model.  
Comparing this figure with Fig. \ref{DOS_J2} and \ref{DOS_J5} it is clear that   $T_{c}$ increases as impurity band forms and 
separates from the edge of the valence band.
For each value of $V$ 
we can identify two values of  $J_c$ for which the slope of the $T_{c}$ vs. $J_c$ curve changes. 
For  $J_c<J_{min}$, 
$T_{c}$ increases very slowly, for $J_{min}<J_c<J_{sat}$ the impurity band begins to develop and $T_{c}$  
increases with the largest slope, for  $J_c>J_{sat}$ the impurity band is completely split from the valence band
and the rate of increase in  $T_{c}$ reduces dramatically. 
In brief, the appearance 
of the impurity band corresponds to the large change in the curvature of $T_{c}$ vs. $J_{c}$.
After the impurity band is well formed increasing  $J_{c}$ or $V$ does not change $T_{c}$ significantly. 
In fact, for $J_c >4\,t$ we 
can anticipate the saturation of the critical temperature. This is an artifact of the DMFA  and is due to 
the absence of non-local correlations. Inclusion of those correlations leads to magnetic frustration of the system, 
which in turn suppresses $T_{c}$.\cite{u_yu_10,g_zarand_02a}
We will come back to this 
point in more detail later when we discuss the two-band model. 

\begin{figure}[tb]
\begin{centering}
\includegraphics[width=0.45\textwidth]{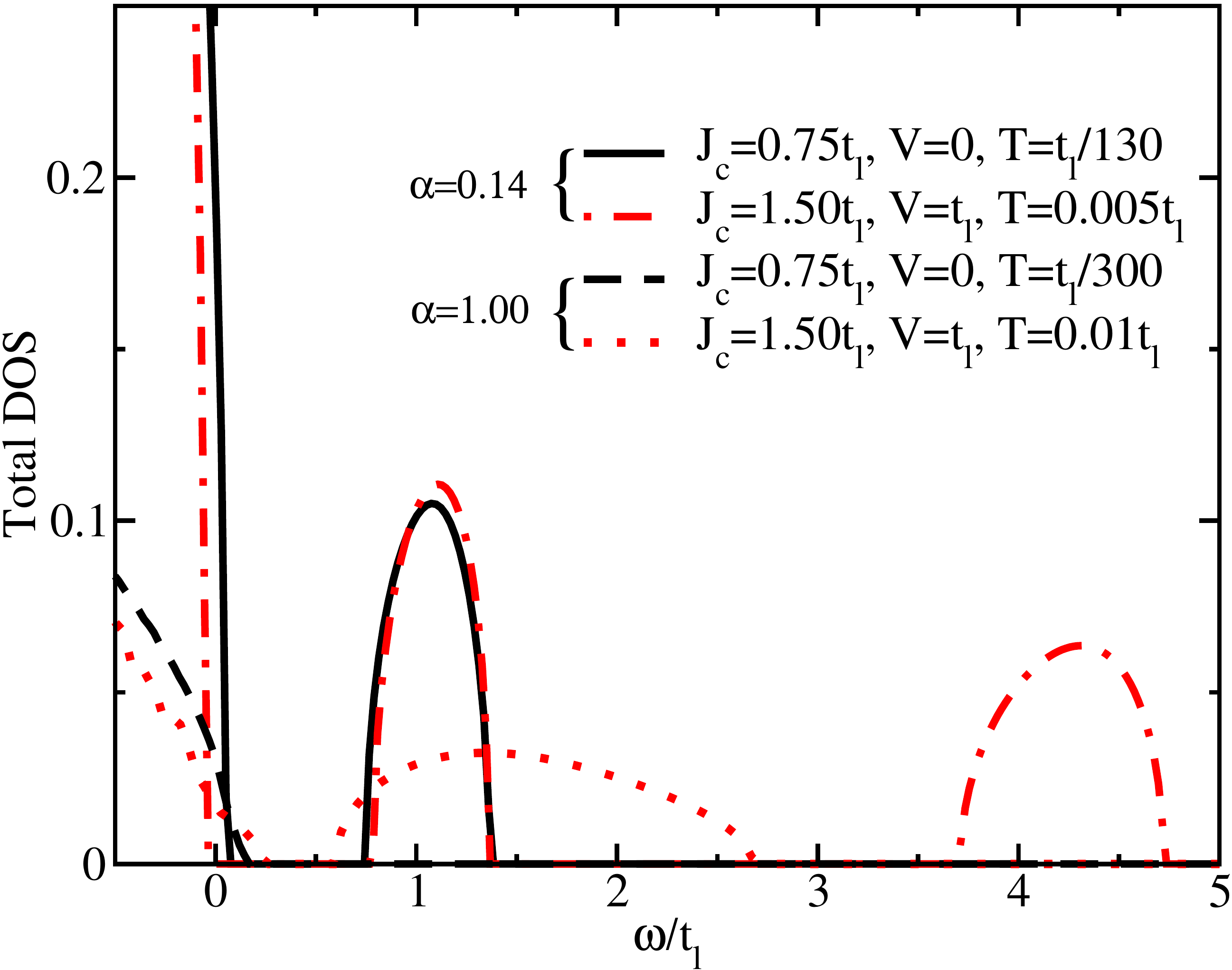} 
\caption{(color online). Density of states 
for two values of the exchange coupling and Coulomb term:
$J_{c}= 0.75\,t_{l}$, $V=0$ and $J_{c}= 1.5\,t_{l}$, $V=t_{l}$ and temperatures well below the 
ferromagnetic transition, and for $\alpha$=0.14 and 1.0. } 
\label{DOS_SO}
\end{centering}
\end{figure}

Therefore by increasing the attractive Coulomb potential $T_{c}$ is significantly enhanced for 
values of the exchange in a given interval, $J_{min}(V)<J_c<J_{sat}(V)$, where  $J_{min}(V)$ and
$J_{sat}(V)$ are function of $V$. This is due mostly to the fact that a positive $V$ promote 
the appearance of localized states at the magnetic sites which mediate the magnetic order.
However, the physics of the ferromagnetic state is not modified by $V$, since the only relevant 
energy scale is given by $T_{c}$, as one expects from a mean field theory. 
This is illustrated in the inset of Fig.~\ref{Tc_vs_J} that displays the polarization of the holes as 
function of $T/T_c$ for a wide range of values of $J_c$ and $V$, showing that all the polarization data 
collapse on a single curve. Thus, the effect of $V$ is just to change the nominal value of $J_c$ to a larger
$J^{eff}_c$.

Now, we introduce a more realistic approach using a two-band model. The spin-orbit interaction and the 
crystal fields lift the 
degeneracy of the $p$-like valence bands into heavy, light and split-off bands. In our 
model we ignore the effect of the split-off band and focus on the heavy and light bands which are 
degenerate at the center of the Brillouin zone.\cite{Luttinger55} $H_{0}$ is approximated by   
$\displaystyle H_0(k)=\hat{R}^{\dag}(\hat{k}) \hat{\epsilon}(k) \hat{R}(\hat{k})\,$, 
where $\hat{\epsilon}(k)$ is a diagonal matrix with entries
$\epsilon(k)_{n,\sigma}= -2 t_n (\rm{cos}(k_x) + \rm{cos}(k_y) + \rm{cos}(k_z))$, with
$n=l,h$ the heavy/light band index
and $\hat{R}(\hat{k})$, the $k\cdot p$ spin $3/2$ rotation matrices.\cite{k_aryanpour_05a}  
In GaAs  the mass ratio of light and heavy holes at the 
$\Gamma$ point is $\alpha$=$m_{l}/m_{h}$= 0.14 \cite{p_yu_01a}. We compare the results of our 
simulation for  $\alpha$=0.14 and $\alpha$=1, keeping the bandwidth of the light hole band fixed.
Furthermore we scale every parameter according to the light holes hopping energy ($t_{l}$), which set
the bandwith of the hole band.

Fig.~\ref{DOS_V0} displays the hole density of states close to the edge of the valence band for 
$J_{c}$= 0.5\,$t_{l}$ and 0.75\,$t_{l}$ in  absence of the Coulomb potential, and for temperatures 
well below 
the ferromagnetic transition temperature. One can anticipate that  
the formation and splitting of the impurity band happens for smaller values of $J_{c}/t_{l}$
than in the one-band model. We can explain this by noting that the total angular momentum of 
the holes can be as large as $J$=3/2 for heavy holes, leading to a larger contribution to the 
total energy from the the second term in Eq.~(\ref{hamiltonian}).  
Moreover, for a small filling 
there are more available states close to the center of the Brillouin zone in the two-band model 
than  in the one-band model. Larger number of spin states available to align along the direction of
the local moment increases the average exchange energy and favor ferromagnetism.

Fig.~\ref{DOS_V1} displays  the density of states for the same exchange couplings and temperatures, 
$J_{c}$= 0.5\,$t_{l}$, T=0.005\,t$_{l}$, and $J_{c}=$0.75\,$t_{l}$, T=1/130\,t$_{l}$, 
but with a finite Coulomb potential $V$=\,$t_{l}$. For these values of the parameters 
a second impurity band appears in the semiconducting gap. 
The appearance of two impurity bands is consistent with the fact that the model includes 
two bands with $J_{z}=\pm 3/2,\pm1/2$. Notice that the second impurity band is more populated with light
holes ($J_{z}=1/2$) while the first impurity band, with higher energy, is mostly made of heavy holes ($J_{z}=3/2$). 
Since we keep the filling of the holes fixed ($n_{h}$=$x/2$) 
the chemical potential sits in the middle of the first impurity band, as shown in Fig. \ref{DOS_V1}. 
Thus, as we discussed previoulsy, the shift of the impurity band
will not have noticeable effects on the magnetic properties of the DMS.

To investigate the effect of the spin-orbit interaction we introduce a simple toy model which has all the features 
of our two-band model except that the heavy and light bands are degenerate over the whole Brillouin zone. 
Therefore, heavy and light bands have the same dispersion but different total angular momenta 
$j_{z}$=$\pm$3/2 and $\pm$1/2, respectively. 
The different band masses introduce magnetic frustration\cite{g_zarand_02a,j_moreno_06a} 
and by setting $\alpha$=1.0 
($m_{h}$=$m_{l}$) in our model, this magnetic frustration is removed.
Since $t_{l}$ is fixed, changes in $\alpha$ alters the dispersion of the heavy hole band while 
keeping the light band fixed.

Fig.~\ref{DOS_SO} displays the total DOS for two values of the exchange coupling and Coulomb potential:
$J_{c}= 0.75\,t_{l}$, $V=0$ and $J_{c}= 1.5\,t_{l}$, $V=t_{l}$, and for $\alpha$=0.14 and 1.0. 
Note that for $\alpha$=0.14 the impurity band is formed at lower couplings.
Thus, the spin-orbit interaction enhances the formation of the impurity band.
We can explain this by noting that changing 
$\alpha$ from 1.0 to 0.14 decreases the kinetic energy of the heavy holes (with $j_{z}$=3/2) becoming
more susceptible to align their spin parallel to the local moment promoting the formation of the 
impurity band. Fig.~\ref{DOS_V0} and \ref{DOS_V1}
show explictly that the heavy holes are the majority of the carriers in the impurity band.
On the other hand the bandwidth of the impurity band is larger 
when $\alpha$=1.0, pointing to less localized holes, which better mediate the exchange interaction
between magnetic ions. 

\begin{figure}[t]
\begin{center}
\includegraphics[width=.45\textwidth]{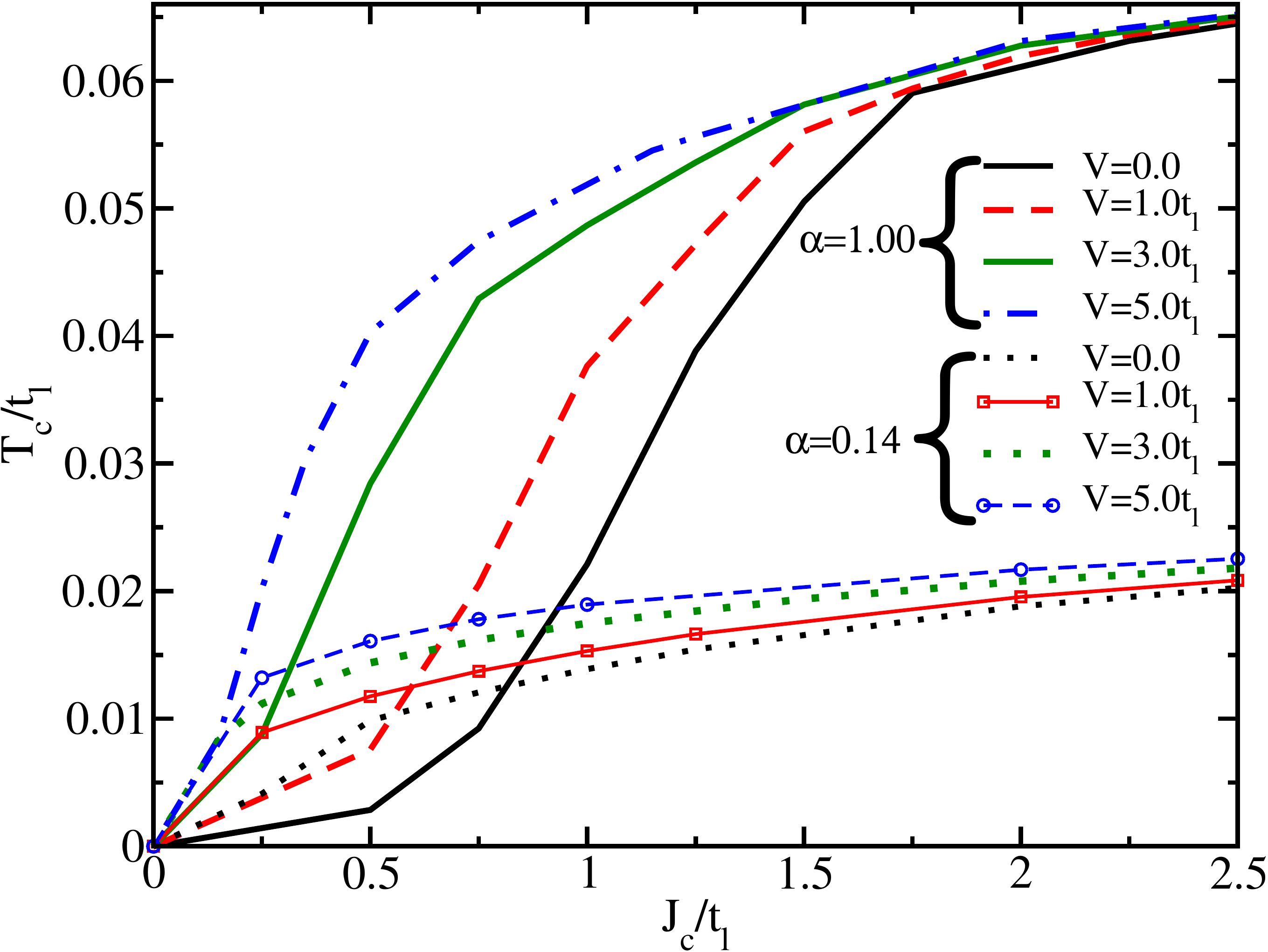}
\caption{(color online). Ferromagnetic transition temperature, 
$T_{c}$, vs. exchange coupling $J_{c}$, both in units of $t_{l}$, for different values of the
Coulomb potential and for $\alpha$=0.14 and 1.0. }
\label{Tc_compare}
\end{center}
\end{figure}

Finally we look at the dependence of the critical temperature on the parameters of the model: $J_{c}$, 
$V$ and $\alpha$. 
The results for different values of $J_{c}$-$V$ for $\alpha$=1.0 and 0.14 are shown in Fig.~\ref{Tc_compare}. 
Similarly to Fig.~\ref{Tc_vs_J} we can identify for both values of  $\alpha$  
a range of parameters $J_{c}$, $V$ where $T_{c}$ increases strongly. This corresponds to the formation and splitting
of the impurity band from the valence band. 
For small values of $J_{c}$ and $V$, $T_{c}$ is higher for $\alpha=0.14$  
but as we increase $J_{c}$-$V$ the ferromagnetic transition temperature 
for $\alpha=1.0$ becomes larger.
Eventually $T_{c}$ saturates due to the lack of non-local correlations within the DMFA. 
We can understand the higher $T_{c}$ for $\alpha=0.14$ and small $J_{c}$,$V$ by looking at Fig.~\ref{DOS_SO}. 
For $\alpha=0.14$ the impurity band appears at smaller values of $J_{c}$ and $V$ than for $\alpha=1.0$.
This is due to the fact that the heavy holes have a smaller kinetic energy and can be polarized more 
easily and become bonded to the localized moments forming the impurity band.
For larger values of $J_{c}$ and $V$, $J_{c}> 0.81 t_{l}$ for $V=0$, $J_{c}> 0.60 t_{l}$ for $V=1\,t_{l}$
or $J_{c}> 0.29 t_{l}$ for $V=3\,t_{l}$, the critical temperature for the model with $\alpha$=1.0 
surpasses the one for $\alpha$=0.14 in agreement with previous findings in the 
strong coupling regime\cite{k_aryanpour_05a,j_moreno_06a}. 
This also can be related with the DOS in Fig.~\ref{DOS_SO}, where the bandwith of the impurity band for $\alpha$=1.0
is larger than for $\alpha$=0.14. A larger bandwidth corresponds to weaker localization of the holes and 
higher mobility. Therefore, they will better 
mediate the ferromagnetic interaction between the magnetic ions and  we expect to see higher $T_{c}$ 
when $\alpha=1.0$. For the largest value of $J_{c}$ and $V$ we study $T_c(\alpha=0.14)/T_c(\alpha=1.)=0.35$
to compare with $0.48$ obtained in the strong coupling limit\cite{j_moreno_06a}. \\

\section{Conclusions}

In conclusion, we have calculated densities of states, polarizations and ferromagnetic transition 
temperatures for a one-band and two-band models appropriate for Ga$_{1-x}$Mn$_{x}$As. 
We have investigated the effect of adding a local Coulomb 
attractive potential $V$ between the magnetic ions and the charge carriers.
The inclusion of a Coulomb term 
leads to the formation of the impurity band for smaller magnetic couplings ($J_{c}$), in agreement 
with previous studies\cite{m_takahashi_03a,f_popescu_07a} and it
significantly enhances $T_c$ for a wide range of $J_c$, without affecting the 
intrinsic physics of the ferromagnetic transition. 
We also explore the effect of the spin-orbit interaction by using a two-band model and two different 
values of the ratio of the effective masses of the heavy and light holes.
We show that in the regime of small 
$J_{c}$-$V$ the spin-orbit interaction enhances $T_{c}$, while for large enough values of $J_{c}$-$V$
the magnetic frustration induced by the spin-orbit coupling reduces $T_c$ to values comparable to 
the previously calculated strong coupling limit.

\vspace{0.3in}

We acknowledge useful conversation with Randy Fishman and Unjong Yu.
This work was supported by the National Science Foundation through
OISE-0730290 and DMR-0548011.
Computation was carried out at the University of North
Dakota Computational Research Center, supported by EPS-0132289 and
EPS-0447679.

\end{document}